\documentclass{iau307}
\usepackage{graphicx}
\usepackage{natbib}
\usepackage{url}
\bibpunct{(}{)}{;}{a}{}{,}

\def\Teff {$T_{\rm eff}$\,}
\def\logg {$\log g$\,}
\def\Msun {$M_\odot$\,}
\def\Mspec {$M_{\rm spec}$\,}
\def\Mevol {$M^{t}_{\rm evol}$\,}
\def\Minit {$M^{init}_{\rm evol}$\,}
\def\Mdot {$\dot M$\,}
\def\logl {$\log L/L_\odot$\,}
\def\vcrit {$v_{crit}$\,} 
\def\vini {$v_{init}$\,}
\def\vsini {$v \sin i$\,}

\title[Mass discrepancy in O stars ] 
{The mass discrepancy problem in O stars of solar metallicity. Does it still exist?}

\author[N. Markova and J. Puls]   
{N. Markova$^1$
 \and J. Puls$^2$}

\affiliation{$^1$IANAO, Sofia, Bulgaria \\ email: {\tt nmarkova@astro.bas.bg} \\[\affilskip]
$^2$Universit\"{a}ts-Sternwarte,	M\"unchen, Germany}

\pubyear{2014}
\volume{307} 
\pagerange{}
\jname{New windows on massive stars: asteroseismology, interferometry, and spectropolarimetry}
\editors{G. Meynet, C. Georgy, J.H. Groh \& Ph. Stee, eds.}

\begin{document}

\maketitle

\begin{abstract}
Using own and literature data for a large sample of O stars in the Milky Way, 
we investigate the correspondence between their spectroscopic and evolutionary 
masses,  and try to put constraints on various parameters that might influence 
the estimates of these two quantities.

\end{abstract}

\firstsection 

\section{Introduction}

In its classical form, the so-called {\it mass discrepancy} refers to the systematic 
overestimate of evolutionary masses, \Mevol, compared to spectroscopically derived 
masses, \Mspec  (e.g., \citealt{herrero92}). While continuous improvements in model 
atmospheres and model evolutionary calculations have reduced the size of the 
discrepancy (e.g., \citealt{repo}), however without eliminating it completely 
\citep{mokiem07, hohle10, massey12}, there are also studies (e.g., \citealt{WV10}) 
which argue that, at least for O stars in the Milky Way, the mass discrepancy problem 
has been solved. 
     
\section{Stellar sample and methodology\label{SecOne}}

 Our sample consists of 51  Galactic dwarfs, giants and supergiants, with 
 spectral types ranging from O3 to O9.7.  Forty one of these are 
 cluster/association members; the rest are field stars. For 31 of the 
 sample stars, we used own determinations of stellar parameters, obtained 
 by means of the latest version of the FASTWIND code (Markova et al., in 
 preparation); for the remaining 20, similar data have been derived by 
 \citet{bouret12} and \citet{martins12a, martins12b}, employing the CMFGEN 
 code instead. 
 
 For all sample stars, \Mspec\,  were calculated  from the effective gravities 
 corrected for centrifugal acceleration, whilst \Mevol\, were determined by 
 interpolation between available tracks along isochrones, as calculated by  
 \citet{ekstroem} (Fig1, upper panels) and \citet{brott} (Fig.1, lower panels). 
 To put constraints on biases originating from uncertain distances and 
 reddening, in parallel to the classical \logl -- log~\Teff diagram 
 (Fig. 1, left panels) we also consider a (modified) spectroscopic HRD 
 (sHRD, Fig. 1, right panels) ) that is independent of 'observed' stellar 
 radii   (for more information, see \citealt{markova14} and \citealt{LK14}).
\begin{figure*}[]
\begin{center}
{\includegraphics[width=6.5cm,height=5.5cm]{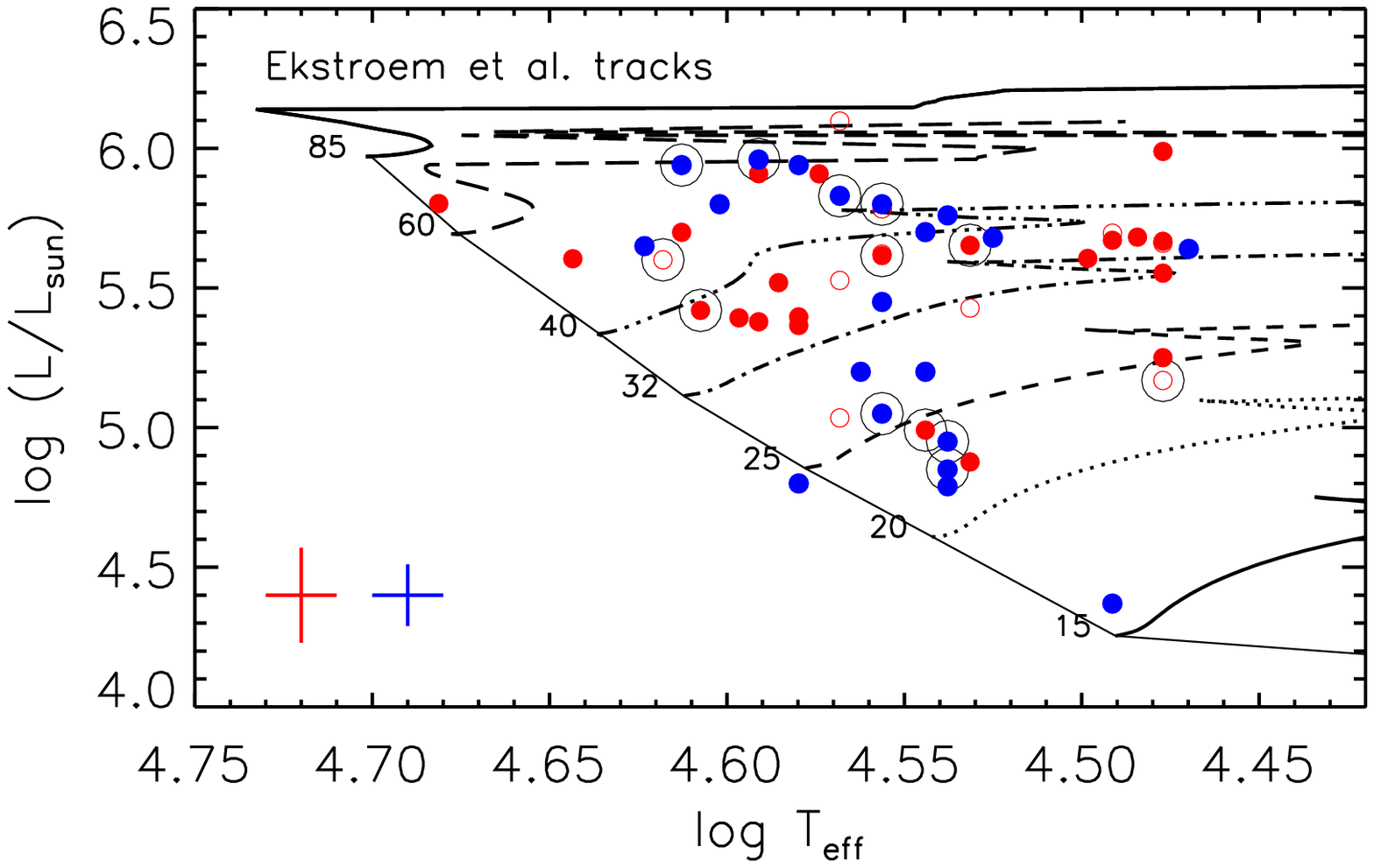}}
{\includegraphics[width=6.5cm,height=5.5cm]{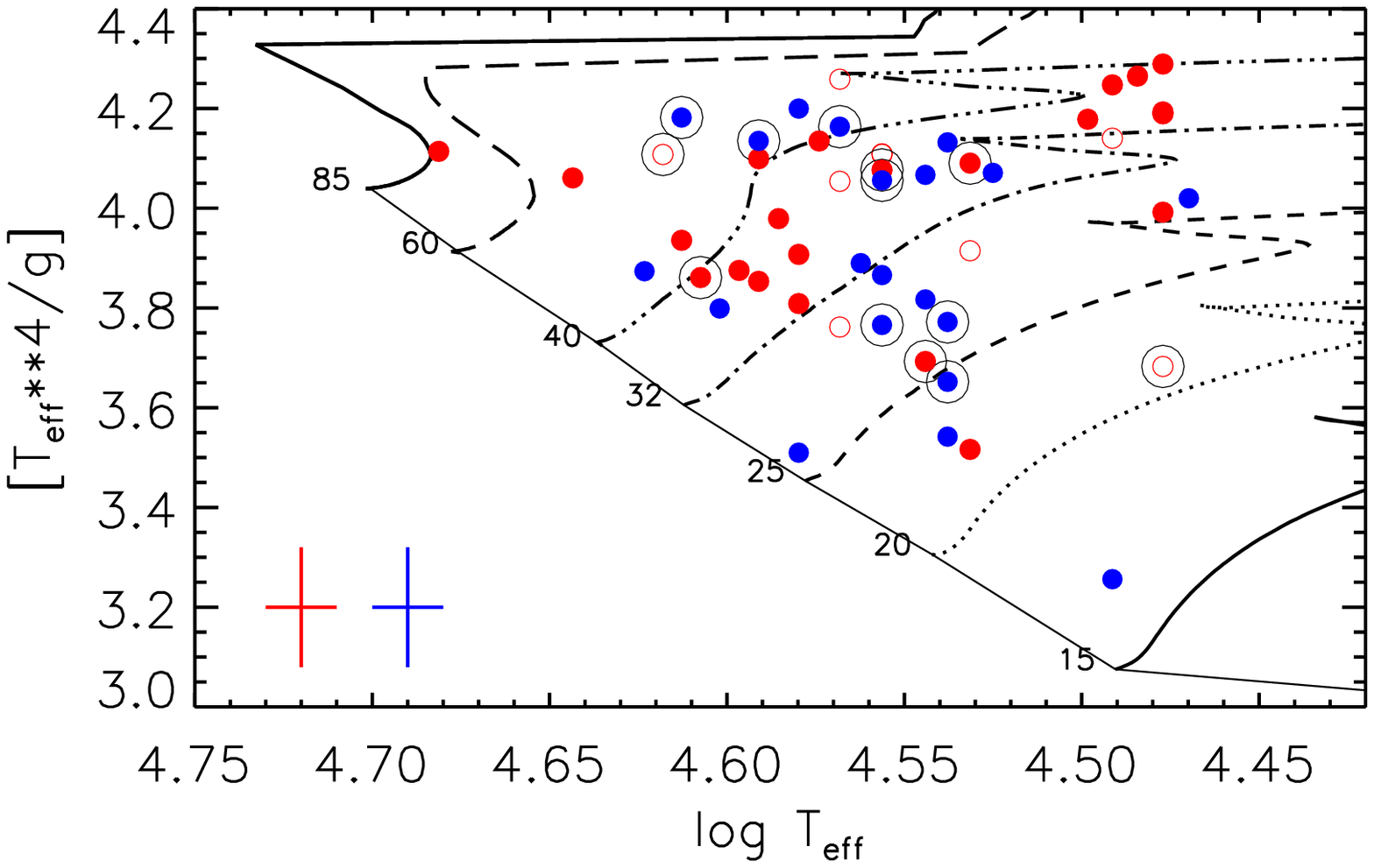}}
{\includegraphics[width=6.5cm,height=5.5cm]{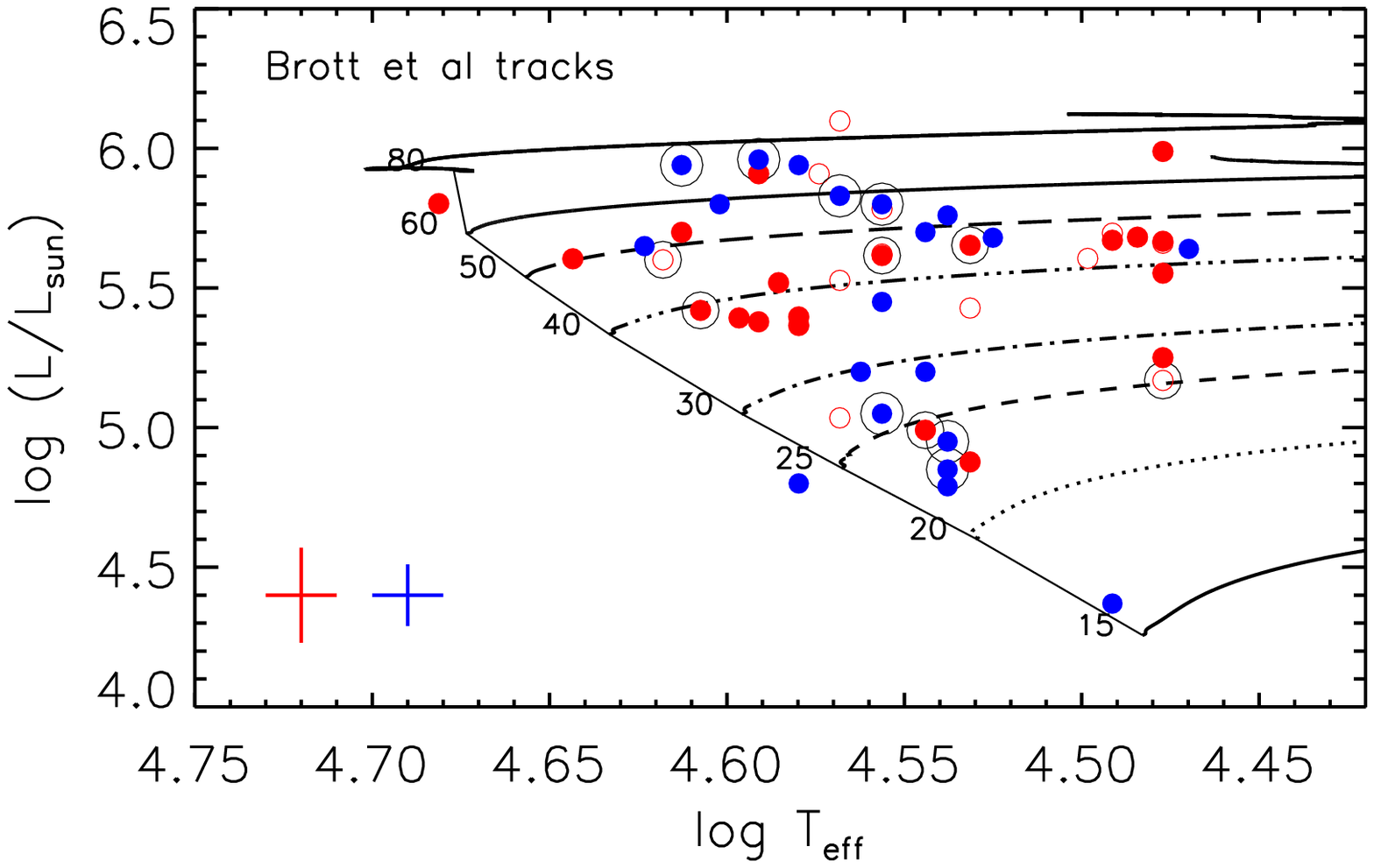}}
{\includegraphics[width=6.5cm,height=5.5cm]{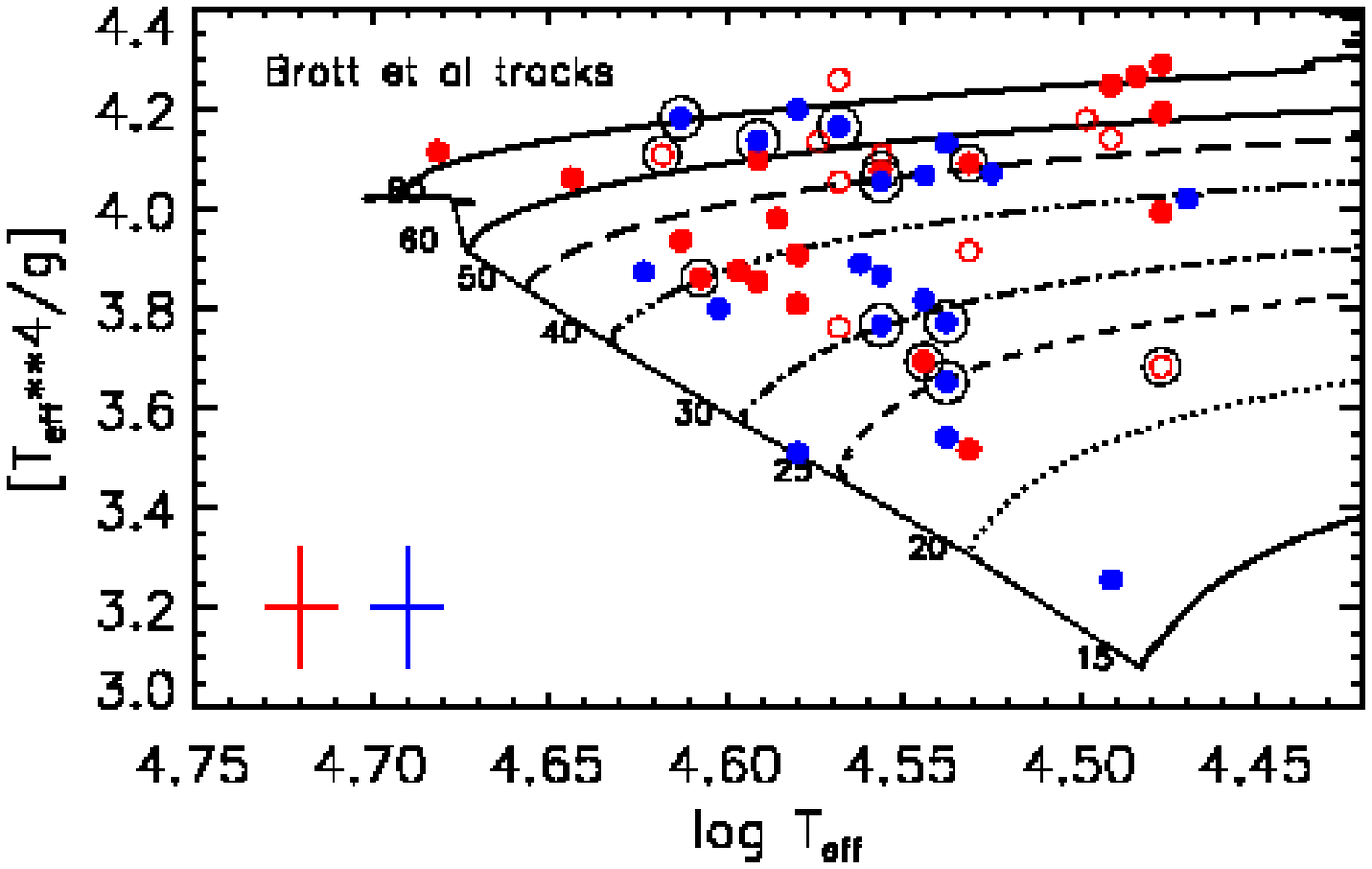}}
\caption{Classical (left) and spectroscopic  (right) HR diagrams,  resulting 
from  Ekstroem et al.  (upper panels) and  Brott et al. (lower panels) tracks 
calculated with \vini  = 0.4 \vcrit  and \vini  = 300 km/s,  respectively, 
compared to own and complementary 'observed' data.  
\newline
{Legend}: red --  FASTWIND data; blue -- CMFGEN data; filled dots -- cluster and 
association members; small open circles - field stars; large open circles - fast 
rotators  (\vsini $>$ 110 km/s, see Markova et al. 2014)
} 
\label{fig1}
\end{center}
\end{figure*}

\newpage
\section{Results}

\begin{figure*}
{\includegraphics[width=6.2cm,height=5.5cm]{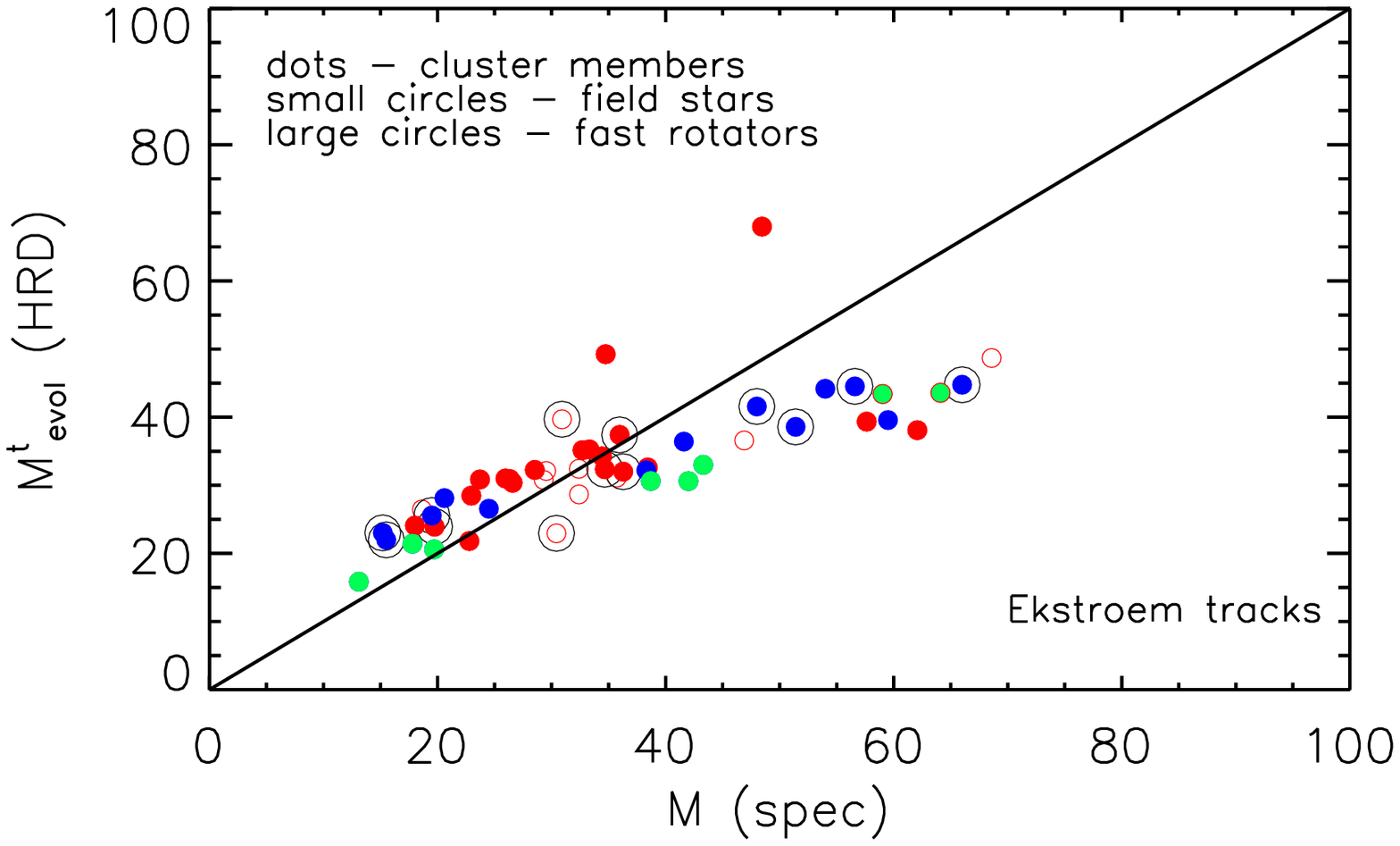}}
{\includegraphics[width=6.2cm,height=5.5cm]{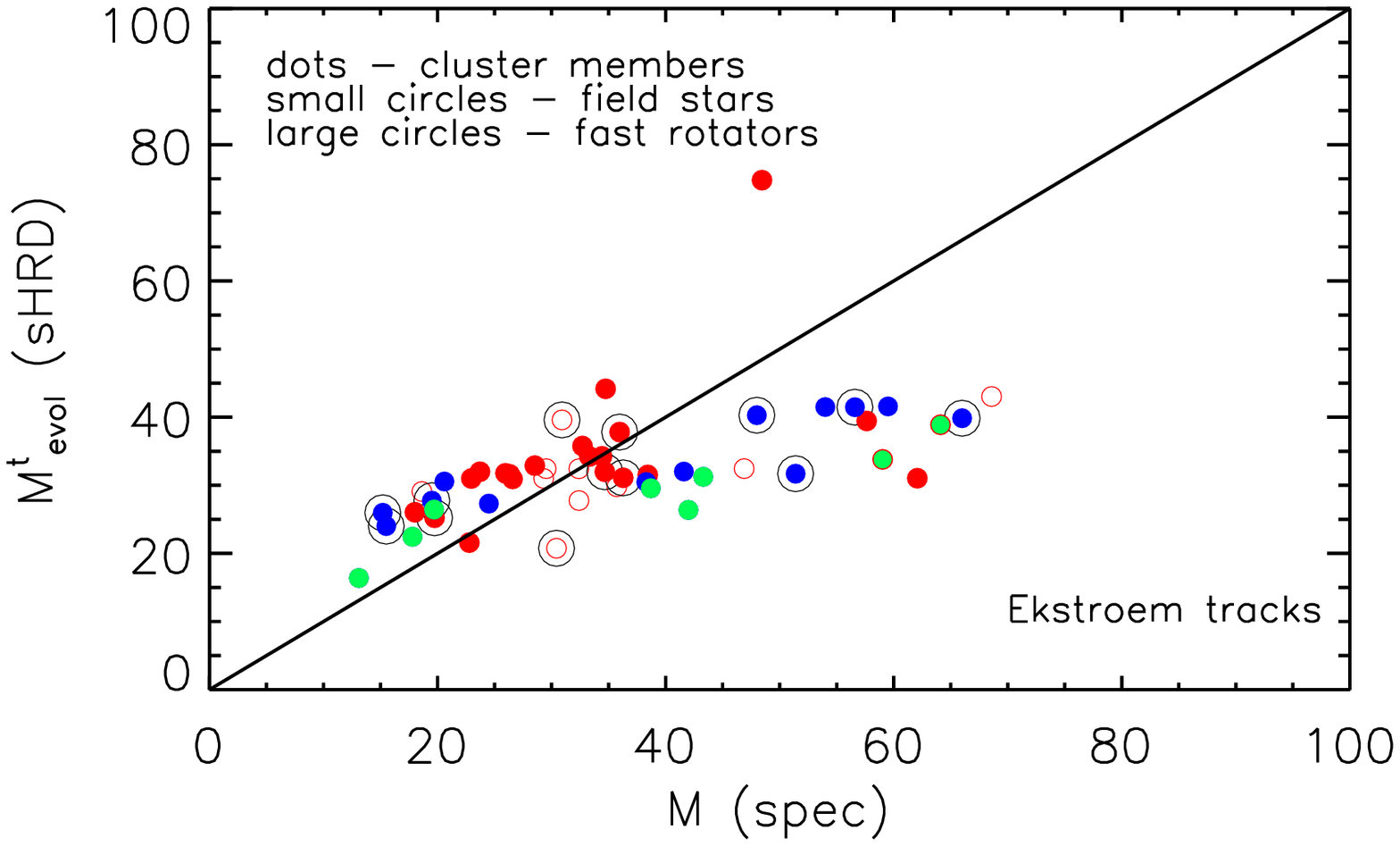}}\\
{\includegraphics[width=6.2cm,height=5.5cm]{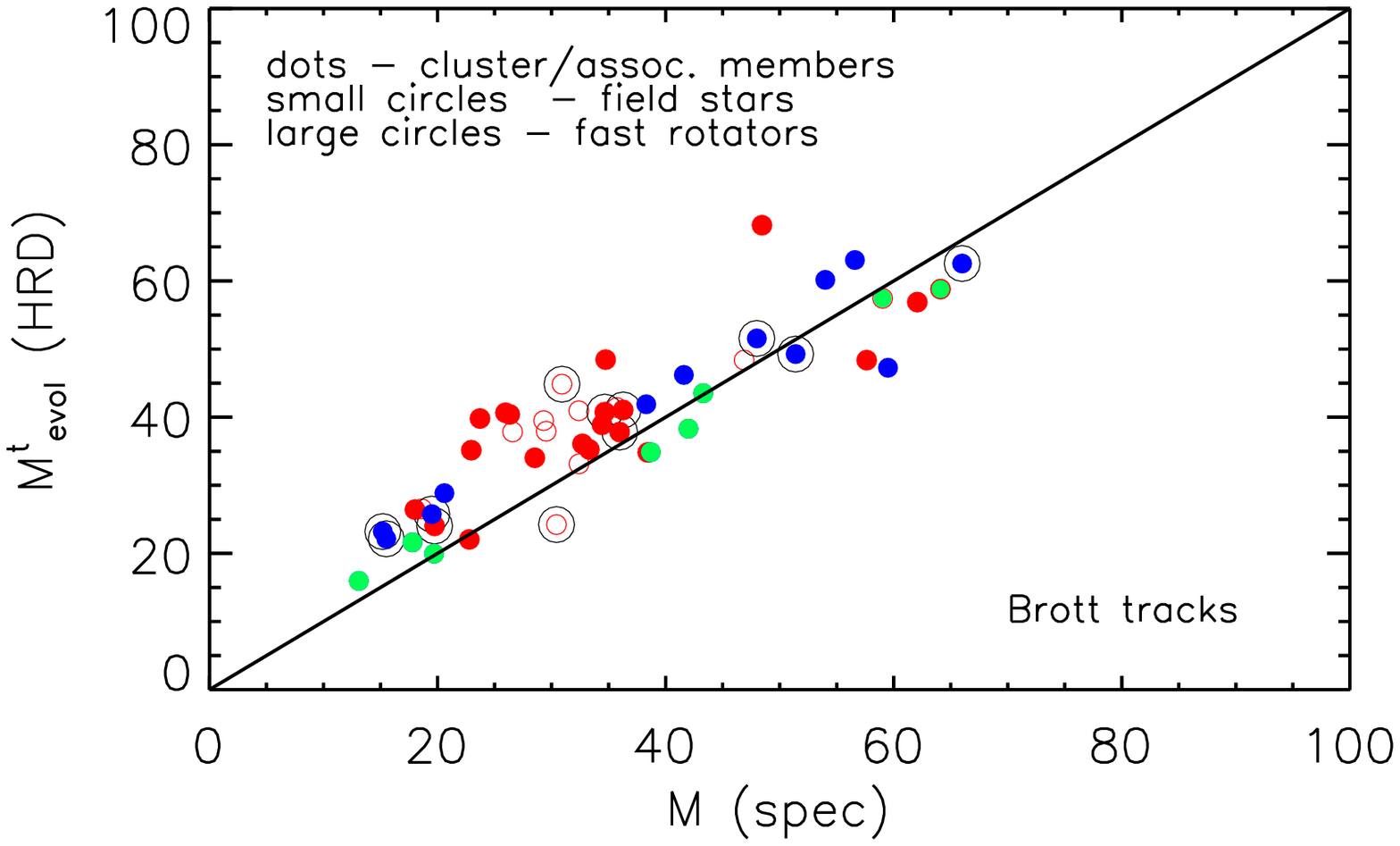}}
{\includegraphics[width=6.2cm,height=5.5cm]{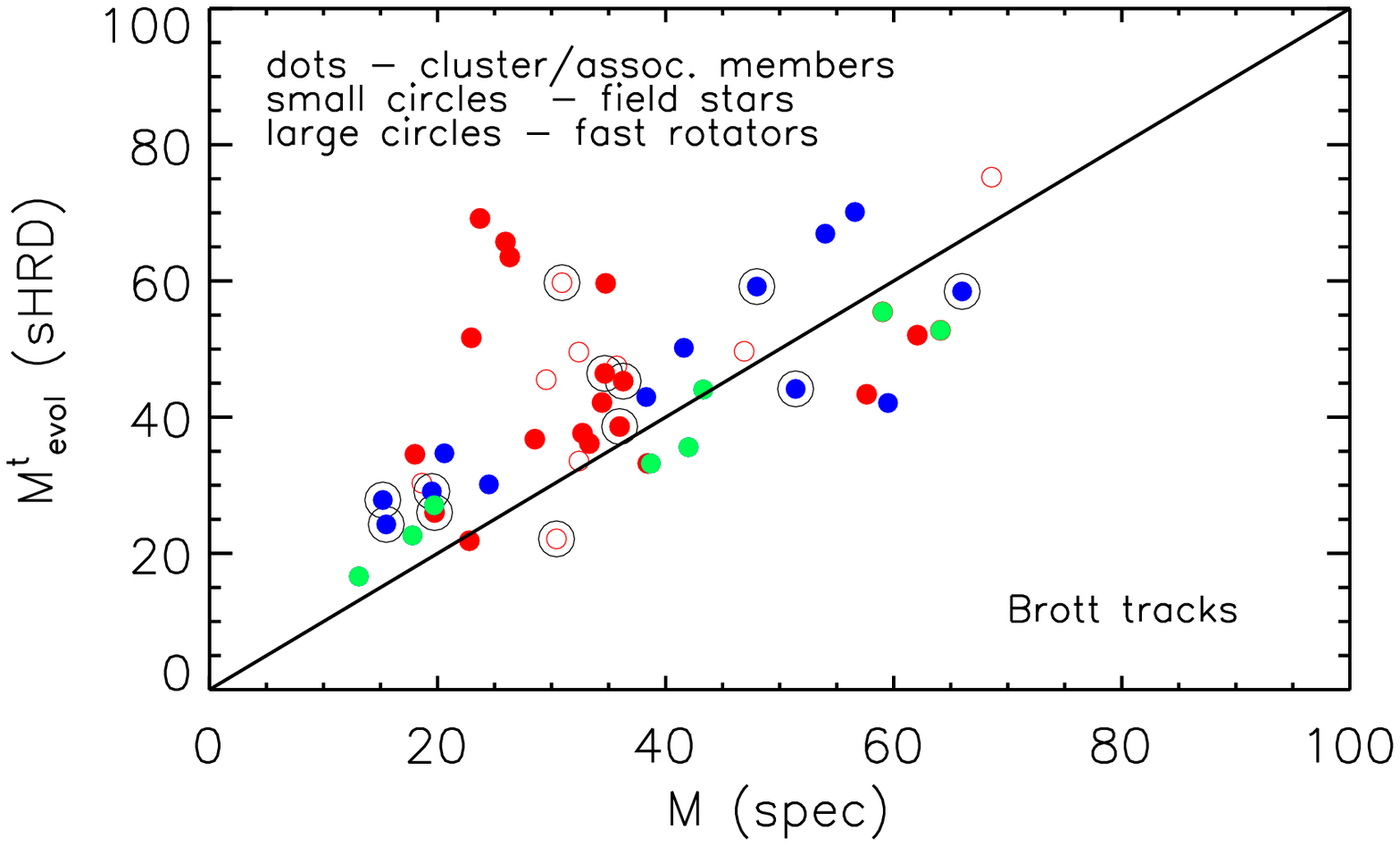}}
\caption{Comparison of \Mevol and \Mspec, for the cases of Ekstroem et al.
(upper panels) and Brott et al. (lower panels) evolutionary models with 
rotation. Although for most of the sample stars  the error bars (not 
indicated) cross the 1:1 line, there is a clear tendency  for  the less 
massive objects (\Mevol below ~35 ..40 \Msun) to show \Mevol $>$ \Mspec. 
Regarding the more massive objects, the derived \Mevol are either 
systematically lower (Ekstroem models) or roughly consistent 
(Brott models) with \Mspec.} 
\label{fig2}
\end{figure*}
Our analysis indicates (see Fig.2) that 
\begin{itemize}
\item[i)] for objects with \Minit $>$ 35 \Msun, \Mevol\, are either systematically 
lower (Ekstroem models) or roughly consistent (Brott models) with \Mspec. As \Mdot 
scales with \logl (e.g., \citealt{vink00}; see also Puls et al., this Volume), 
and as -- soon after the ZAMS -- the 
Ekstroem models with rotation and \Minit$\ge$40 \Msun  become more luminous than 
the Brott models of the same \Minit and \Teff,  we suggest that the $negative$ mass 
discrepancy established for the Ekstroem tracks is most likely related  to 
(unrealistically?) high mass-loss rates implemented in these models. (Warning! The 
good agreement between \Mspec and \Mevol read off the Brott tracks does not 
necessarily mean that the corresponding mass-loss rates are of the right order of 
magnitude, see next item)
\item[ii)] for objects with \Minit$<$ 35\Msun, \Mevol tend to be larger than \Mspec. 
As massive hot stars can develop  subsurface convection zones \citep{cantiello09}, 
and as they can be also subject to various instabilities, we are tempted to speculate 
that  the neglect of turbulent pressure  in FASTWIND and CMFGEN atmospheric models 
might explain the lower \Mspec compared to \Mevol\footnote{By including such a 
turbulent pressure, one would obtain a spectroscopic \logg that is larger by 0.2 dex, 
for typical parameters and a turbulent speed of 15 km/s}. Indeed,  one might argue that 
if our explanation was correct a similar discrepancy should be present (but is not observed) 
for  the more massive stars as well. However, such caveat might be easily solved if also 
the Brott models over-estimate the mass-loss rates, as already suggested by \citet{markova14}, 
and as also implied from up-to-date comparisons of theoretical and observed \Mdot\, (e.g., 
\citealt{najarro11, cohen14})
\item[iii)] while for most sample stars the correspondence between \Mspec  and \Mevol 
does not significantly depend on the origin of the latter (HRD or sHRD), there are a 
number of outliers which, for the case of Brott tracks, demonstrate 
\Mevol(sHRD) $>$  \Mevol (HRD), by  a factor of 1.5 to 1.8. While specific reasons, 
such as, e.g., close binary evolution or homogeneous evolution caused by  rapid rotation, 
can in principle explain discrepant masses read off the HRD and sHRD \citep{LK14},  
it is presently unclear why this discrepancy does not appear in the  Ekstroem tracks.
\item[iv)]the established mass discrepancy does not seem to be significantly biased by 
uncertain stellar radii; the presence of surface magnetic fields, or systematically 
underestimated \logg-values 
derived by means of the FASTWIND code (for more information, see \citealt{massey13}).
\end{itemize}

\acknowledgements{NM and JP gratefully acknowledge a travel grand by
respectively IAU and the University of Geneva. JP also acknowledge 
support by the German DFG under grand PU117/8-1.}

\bibliographystyle{iau307}
\bibliography{MyBiblio}

\end{document}